\documentclass{moriond}

\usepackage{gensymb}
\usepackage{aas_macros}

\def\be{\begin{equation}}
\def\ee{\end{equation}}
\def\bea{\begin{eqnarray}}
\def\eea{\end{eqnarray}}

\newcommand{\Msol}{M_\odot}
\newcommand{\Photo}{\includegraphics[height=30mm]{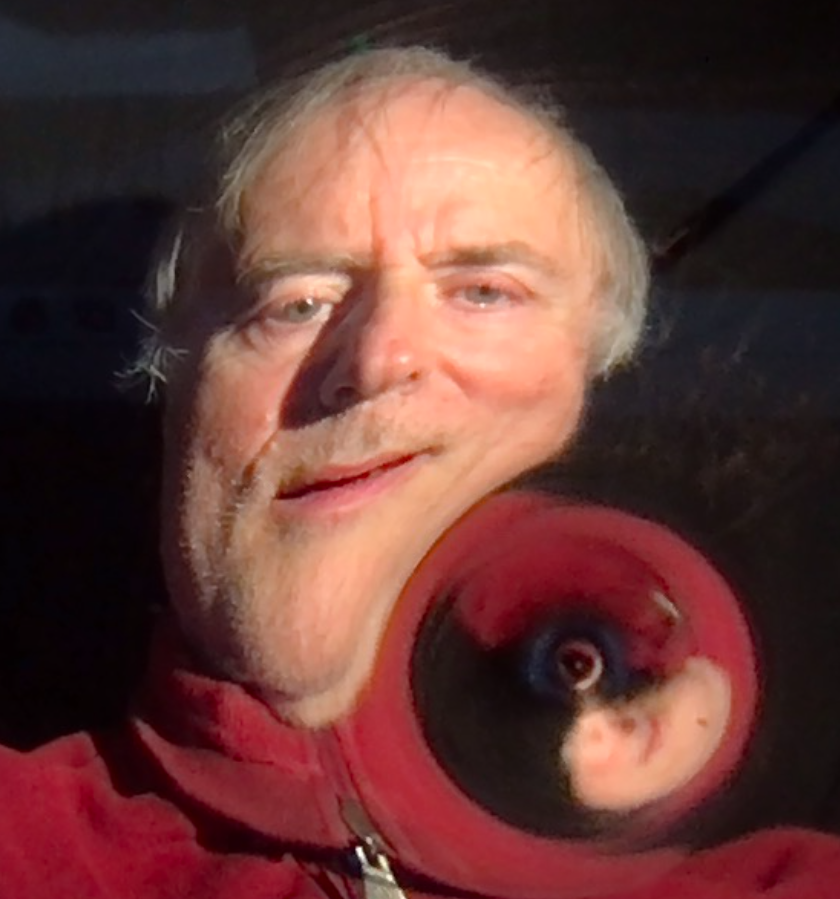}}

\begin{document}
\vspace*{4cm}
\title{Search for Black Holes in the Galactic Halo by Gravitational Microlensing\footnote{This paper uses public domain data obtained by the MACHO Project, jointly funded by the US Department of Energy through the University of California, Lawrence Livermore National Laboratory under contract No. W-7405-Eng-48, by the National Science Foundation through the Center for Particle Astrophysics of the University of California under cooperative agreement AST-8809616, and by the Mount Stromlo and Siding Spring Observatory, part of the Australian National University.}
}

\author{ T. Blaineau, M. Moniez, presented by M. Moniez}

\address{Laboratoire de physique des 2 infinis Ir\`ene Joliot-Curie,
CNRS Universit\'e Paris-Saclay,
B\^at. 100, Facult\'e des sciences, F-91405 Orsay Cedex, France}

\maketitle\abstracts{
Black hole-like objects with mass greater than $10 \Msol$, as discovered by gravitational antennas, can produce long time-scale (several years) gravitational microlensing effects. Considered separately, previous microlensing surveys were insensitive to such events because of their limited duration of 6-7 years.
We combined light curves from the EROS-2 and MACHO surveys to the Large Magellanic Cloud (LMC) to create a joint database for 14.1 million stars, covering a total duration of 10.6 years, with fluxes measured through 4 wide passbands. We searched for multi-year microlensing events in this catalog of extended light curves, complemented by 24.1 million light curves observed by only one of the surveys.
Our analysis, combined with previous analysis from EROS, shows that compact objects with mass between $10^{-7}$ and $200 \Msol$ can not constitute more than $\sim 20\%$  of the total mass of a standard halo (at $95\%$ CL). We also exclude that $\sim 50\%$ of the halo is made of Black Holes (BH) lighter than $1000 \Msol$.}

\section{Introduction: microlensing toward LMC}
\begin{figure}
    \centering
    \includegraphics[width=0.6\linewidth]{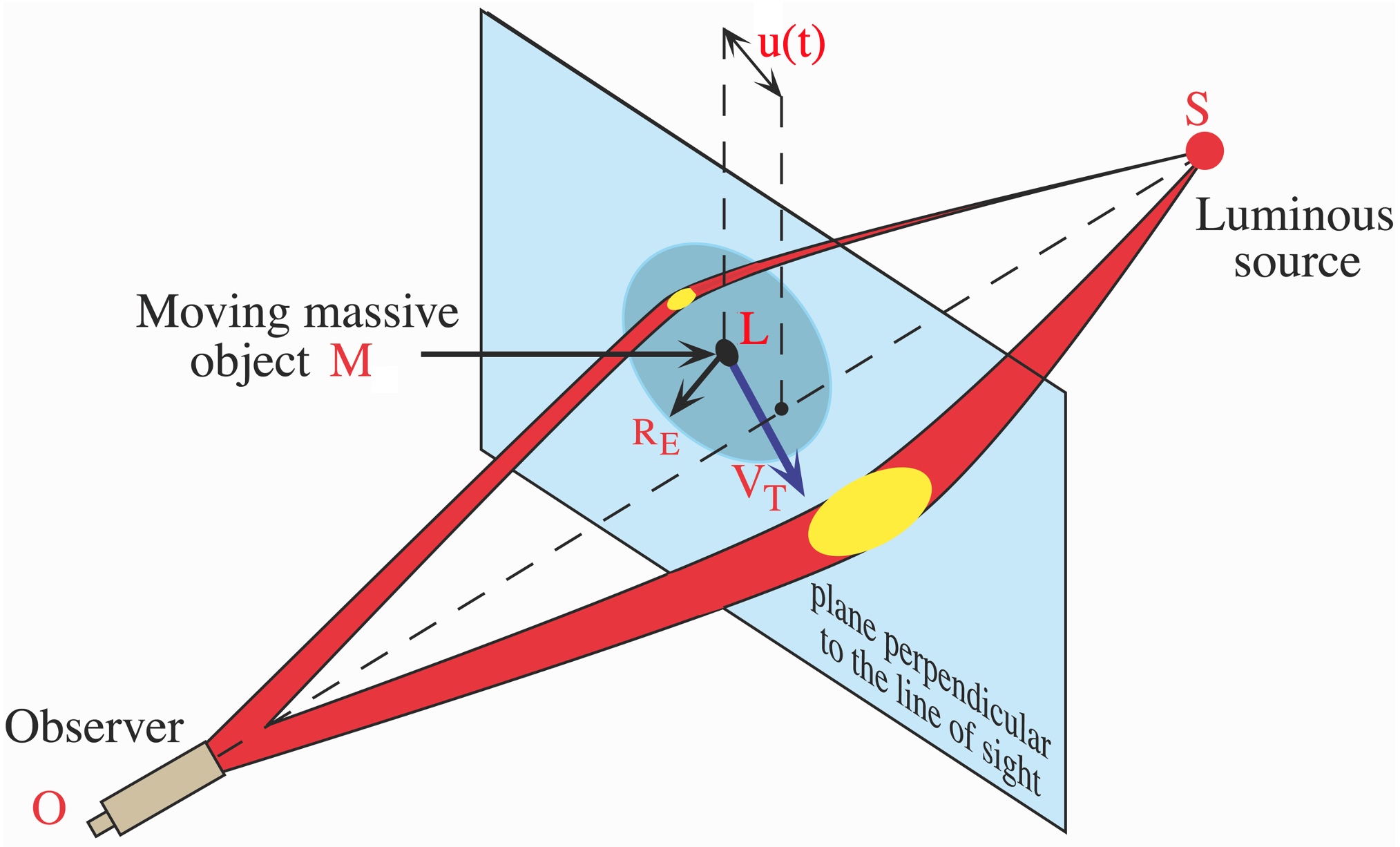}
    \caption{
Principle of the microlensing effect: As the lens (L) of mass $M$ moves with a transverse relative velocity $v_T$ , the impact parameter $u(t)$ changes with time, and so does the magnification of the source.
    }
    \label{fig:principe}
\end{figure}

When a point object (lens) of mass $M$ located at distance $D_L$ from an observer passes
close enough to the line of sight of a point source at distance $D_{S}$,
the observer receives a double image from that source (Fig. \ref{fig:principe}).
These images are not resolved by telescopes but, if they are not occulted by the lens,
the luminosity of the source appears to be temporarily magnified by a time-dependent factor
$A(t)$ according to\cite{Paczynski86}:
\begin{equation}
A(t)=\frac{u(t)^2+2}{u(t)\sqrt{u(t)^2+4}},
\label{Amplification}
\end{equation}
where $u(t)$ is the distance of the lens to the undeflected line of sight, divided by the Einstein radius $r_{\mathrm{E}}$,
\begin{equation}
r_{\mathrm{E}} = \sqrt{\frac{4GM}{c^2}D_{S}.x(1 - x)}
\simeq 4.5\mathrm{AU}\times\left[\frac{M}{\Msol}\right]^{\frac{1}{2}} \left[\frac{D_S}{10\,kpc}\right]^{\frac{1}{2}}
\frac{\left[x(1 - x)\right]^{\frac{1}{2}}}{0.5}.
\end{equation}
$G$ is the Newtonian gravitational constant, $\Msol$ is the mass of the Sun, and $x = D_L/D_{S}$.
The Einstein radius of the lens $r_{\mathrm{E}}$ is such that a point source that is behind the
Einstein disk (of surface $\pi r_{\mathrm{E}}^2$) is magnified by a factor greater than 1.34.

If the lens has a constant relative transverse velocity $v_T$ relative to the line of sight, $u(t)$ is given by
$u(t)=\sqrt{u_0^2+(t-t_0)^2/t_{\mathrm{E}}^2}$,
where $t_{\mathrm{E}}=r_{\mathrm{E}} /v_T$ is the Einstein radius crossing time,
and $u_0$ is the minimum distance to the undeflected line of sight at time $t_0$.

The optical depth $\tau$ up to a given source distance, $D_S$, is defined as the probability
to intercept a deflector's Einstein disk, which corresponds to a magnification $A > 1.34$.
It is found to be independent of the deflectors' mass function:
\begin{equation}
\tau=\frac{4 \pi G D_S^2}{c^2}\int_0^1 x(1-x)\rho(x) dx ,
\end{equation}
where $\rho(x)$ is the mass density of deflectors at distance $x D_S$.

Microlensing has been searched towards the Large Magellanic Cloud (LMC) since the 1990s, to
detect hypothetical massive compact objects in the dark halo.
Assuming that the Milky Way dark matter halo is isotropic and isothermal\cite{Alcock2000} (S-model), and
using its most recent parameters (as in (\cite{Blaineau2022})),
the optical depth to the LMC ($D_{LMC}=49.5\,$kpc) is $\tau_{LMC}\sim4.7\times 10^{-7}$.
If all lenses have the same mass $M$, then 
the mean duration of the events is $\langle t_E\rangle\sim 63\,$days$\times\sqrt{M/\Msol}$.
Past microlensing searches to the LMC have shown that objects with masses $10^{-7}<M<10 M_{\odot}$ do not contribute significantly to the hidden mass of the Milky Way's spherical halo\cite{Tisserand_2007,MACHO_2001,Wyrzykowski_2011}, but the analysis of these surveys was insensitive to the long time-scale events expected from heavier objects such as those responsible for gravitational wave emissions.
To explore the dark matter halo beyond $10 M_{\odot}$ by searching for longer duration events\cite{Mirhosseini2018} (several years), we combined the EROS-2 and MACHO databases\cite{these_Blaineau,Blaineau2022}, which were acquired during different periods.

\section{Combining EROS-2 and MACHO data}
\label{section:combine}
\begin{figure}
    \centering
        \includegraphics[width=0.55\linewidth]{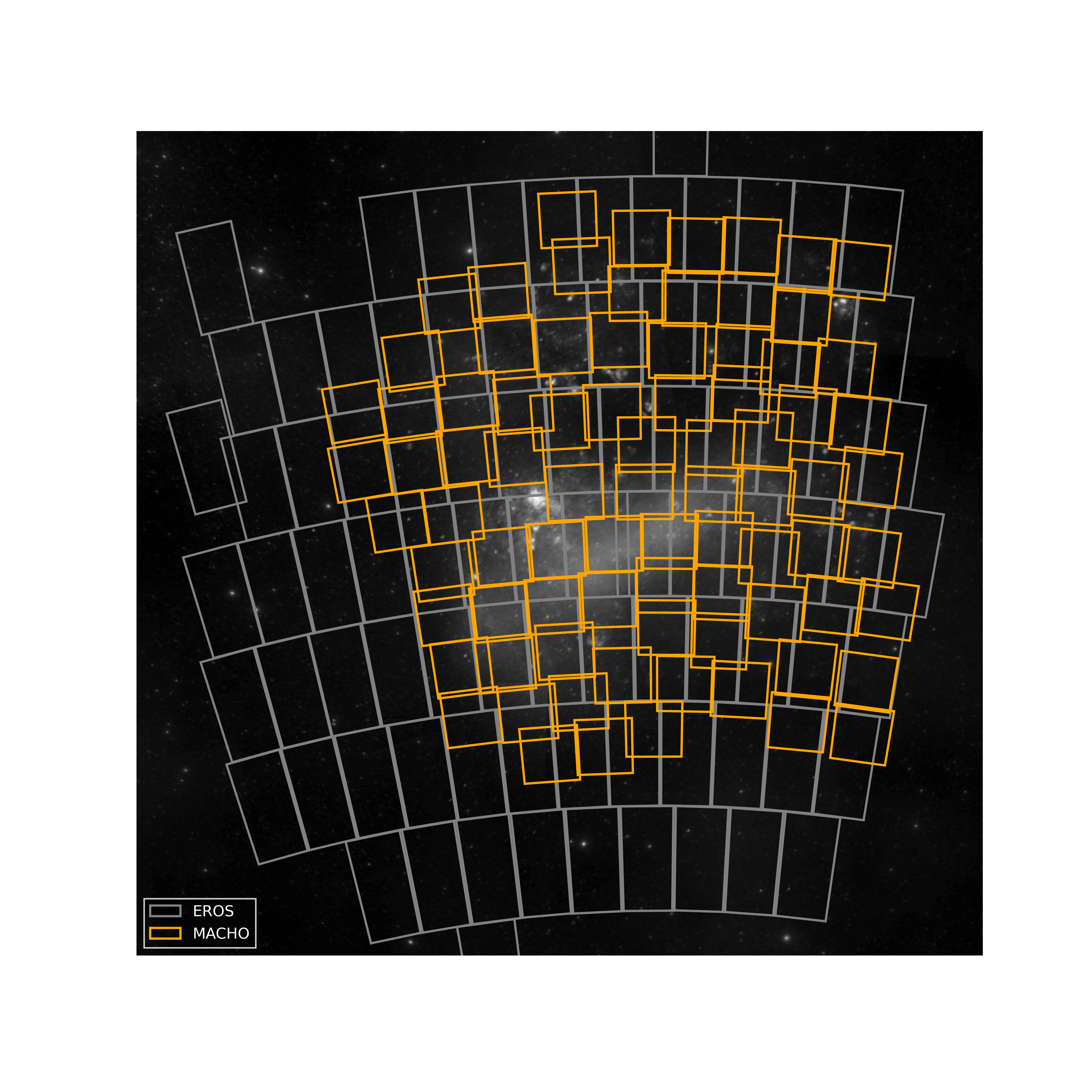}
    \includegraphics[width=0.4\linewidth]{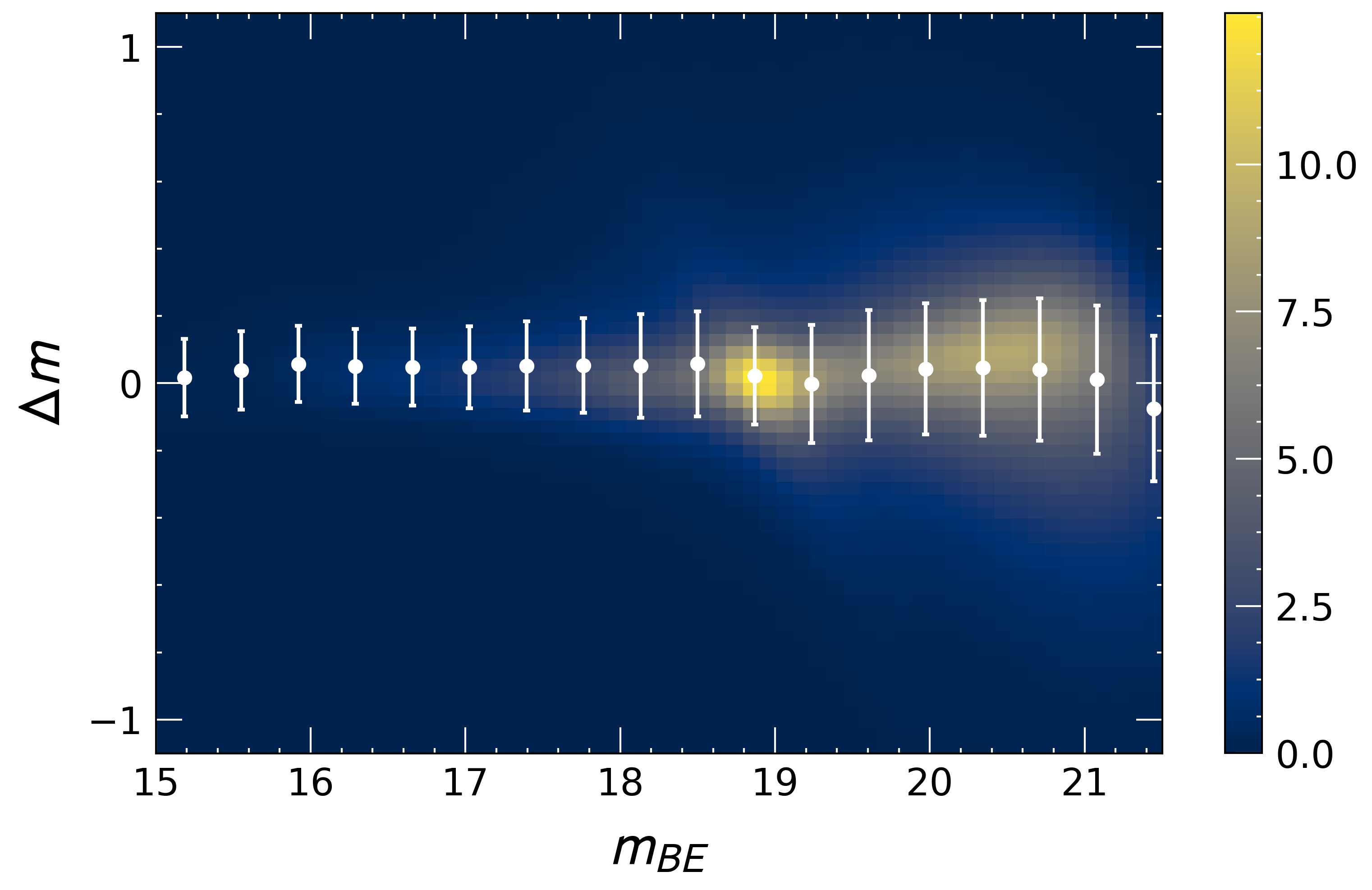}
    \caption{
    Left:
Arrangement of the EROS (grey) and MACHO (orange) fields superimposed on an LMC image. Field size is $\sim 10\degree\times 10\degree$. \\
    Right:
Distribution of the EROS-2 and MACHO magnitude difference $\Delta m=m_{BE}-m_{BM}$ as a function of the EROS-2 blue magnitude $m_{BE}$ for the 14.1 million of associated objects; the values of $m_{BM}$ are derived from the original MACHO magnitudes using a first-degree color equation, so as to match the EROS blue magnitudes on average.
    As a consequence, the mean value of $\Delta m$ varies by a few percent depending on the color of the sources, in particular between the main sequence and the red giants branch (around $m_{BE}=18.8$).
    White dots and bars show the average and standard deviations of $\Delta m$ for each $m_{BE}$ slice. Color scale is in million of objects per squared magnitude.
    }
    \label{fig:assoc}
\end{figure}

The EROS-2 at La Silla Observatory - Chile (MACHO at Mount Stromlo Observatory - Australia) survey setup, consisting of a $1.0m$ ($1.27m$) telescope and two cameras with $8$ ($4$) CCD of 2Kx2K pixels each, monitored 88 (82) fields of 1.deg$^2$ (0.5deg$^2$) towards the LMC (see Table \ref{Table-catalogue} and Fig. \ref{fig:assoc}, left).
The MACHO light curves and images are publicly available\footnote{https://macho.nci.org.au/} \cite{Alcock1999},
and the EROS-2 catalog for LMC was produced for the final EROS LMC publication\cite{Tisserand_2007}.

To associate objects in the two catalogs, available MACHO and EROS-2 sky coordinates $(RA, DEC)$ were refined using {\it Gaia} EDR3 astrometry\cite{2020Gaia_arXiv}, correcting for local shifts up to $2"$ for MACHO and $0.5"$ for EROS-2.
After this correction we were able to associate the objects of the two surveys with a precision of better than $0.1"$.
Considering the typical spread of the light of the stars on the best images ($FWHM\sim 1"$), and the similar resolutions of the surveys, the associated reconstructed objects of each survey contain the same stars, with little variation of the blend components. 
This is confirmed by the good correlation observed between the EROS-2 and MACHO fluxes (Fig. \ref{fig:assoc}, right),
which is compatible with the photometric accuracy.

\begin{table}[h!]
\begin{center}
 \caption[]{Statistics of the MACHO and EROS-2 surveys.
    Survey durations, number of monitored objects, median stellar densities, approximate limiting magnitudes, median numbers of flux measurements per object after cleaning.}
\begin{tabular}{|l|c c c |}\hline
    & EROS-2 only & MACHO only & common \\
    \hline
    Dates (month/yr) & 7/96-2/03 & 7/92-1/00 & 7/92-2/03 \\
    $T_{obs}$ (year) & 6.7 & 7.7 & 10.6 \\
    $N_{objects} (\times10^6)$ & 15.8 & 6.9 & 14.1 \\
    \hline
    \footnotesize{\bf central fields $(\degree)^2$} & $\sim 10$ & $\sim 10$ & $\sim 10$ \\
    $\rm stars/arcmin^2$ & $\sim 70$ & $\sim 100$ & $\sim 70$ \\
    mag. lim. $V_{Cousins}$ & $\sim 20.5$ & $\sim 20.5$ & $\sim 20.5$ \\
    \# measurements B & 500 & 1400 & 1900 \\
   \# measurements R & 600 & 1550 & 2150 \\
    \hline
    \footnotesize{\bf outer fields $(\degree)^2$} & $\sim 77$ & $\sim 39$ & $\sim 39$ \\
    $\rm stars/arcmin^2$ & $\sim 30$ & $\sim 20$ & $\sim 20$ \\
    mag. lim. $V_{Cousins}$ & $\sim 22.5$ & $\sim 21.5$ & $\sim 21.5$ \\
    \# measurements B & 250 & 200 & 450 \\
   \# measurements R & 300 & 250 & 550 \\
   \hline
   \end{tabular}
   \centering
    \label{Table-catalogue}
\end{center}
\end{table}
\section{Event selection}
$14.1\times10^6$ objects benefit from 10.6 years of luminosity measurements, of which 3.8 years overlap, with photometric
series corresponding to the 4 passbands (2 per survey).
$22.7\times10^6$ objects, monitored by only one survey, were also included in our search, although they are monitored for shorter times.
The details of the analysis presented here can be found in (\cite{Blaineau2022}).

We first eliminated measurements associated with the remaining problematic images (blurred
or with a guiding or readout defect), and poor measurements due to instrumental effects.
We then renormalized the photometric uncertainties of EROS-2 and MACHO for each light curve so that the time-averaged normalized uncertainties correspond to the point-to-point flux dispersions along the curve.

\begin{figure}
    \centering
    \includegraphics[width=0.8\linewidth]{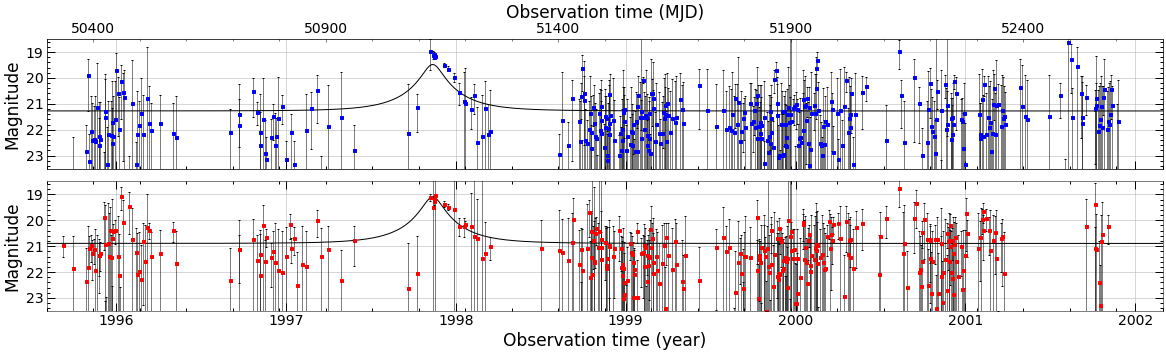} \\
    \includegraphics[width=0.8\linewidth]{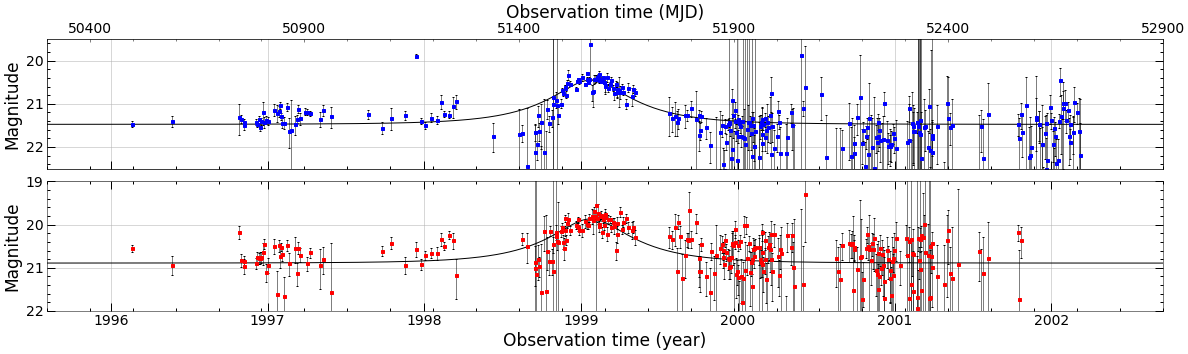}
    \caption{
    Light curves (magnitude vs. time) of candidates {\it lm0690k17399} (upper panels) and {\it lm0073m17729} (lower panels) in EROS blue passband, and EROS red passband. The black solid lines show the best no-blend microlensing fit.
   These objects are not in the MACHO field.
    }
    \label{fig:candidates}
\end{figure}

We performed a discriminant analysis on the light curves of the $36.8\times10^6$ objects, based on comparing the fit of a microlensing event with that of a constant light curve.
For each object we performed a simultaneous point-source point-lens microlensing fit to the available light curves, with one base flux line per passband and a set of microlensing parameters $(t_0,u_0,t_E)$ common to all passbands (fit without blending or parallax).
We then require that light curves with a good microlensing fit have their maximum brightness between $[t_{start}+200\,$days$,t_{end}]$,
where $t_{start}$ and $t_{end}$ are the instants of the start and end of the measurements, and that
$100\,$days$\,< t_E < (t_{end}-t_{start})/2$.
These latter criteria reject most short-lived eruptive events and in particular supernovae.
We eliminate so-called "blue bumpers", Be class stars that sometimes show asymmetric bumps,
by making a stricter selection in their color-magnitude diagram (CMD) region.
After this preselection, we reject three types of objects that show variabilities that can be confused with long-lasting microlensing effects:
Objects outside the region of the CMD containing $99\%$ of the stars, more likely to show variability;
echoes from SN1987A;
and variable objects identified in external catalogs.
Only two microlensing candidates satisfy the whole selection process (Fig. \ref{fig:candidates}).
These candidates are probably not microlensing effects, but they illustrate the sensitivity of our analysis
to long time-scale bumpers:
Candidate {\it lm0690k17399} is probably a type II-L or II-P supernovae, and {\it lm0073m17729}
shows hints of variability outside the main event.
However, we cannot formally exclude these candidates without external data or stricter selection, so
we conservatively keep them in our calculation of limits on the macho content of the Halo, but cannot use them to
derive a microlensing optical depth.
\section{Detection efficiency, expected rate, and limits on heavy objects in the halo}
\label{section:efficiency}
\begin{figure}
    \centering
    \includegraphics[width=0.55\linewidth]{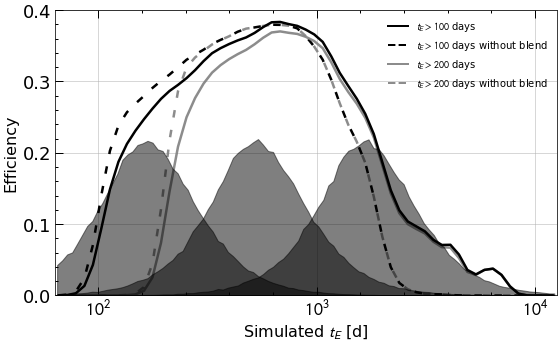}
    \caption{
    Detection efficiency:
    Solid (dashed) lines show efficiencies taking (or not) into account blending effects. 
    Grey lines show efficiencies of the analysis adding the constraint $t_E>200$ days.
Grey histograms (not normalized) are the expected $t_E$ distributions for a halo made of $10$, $100$, and $1000\Msol$ compact objects (from left to right).
    }
    \label{fig:tEdist_eff}
\end{figure}

The detection efficiency $\epsilon(t_E)$ (Fig. \ref{fig:tEdist_eff}), defined as the ratio of the number of events that satisfy
our selection to the number of events with $u_0<1$ and $t_0$ within the observation period,
was computed by subjecting simulated events to our selection procedure.
To be realistic, the simulations were produced by superimposing simulated microlensing
events on the light curves of a representative random subsample of the observed objects.
We took into account ``blending''
by assigning a set of stars to each object in a way that is consistent
with the observed density of LMC stars in Hubble Space Telescope (HST) images.
We also estimated that blending from undetected binarity on HST images
has a negligible impact on the evaluation of the detection efficiency for long duration events.

\begin{figure}
    \centering
    \includegraphics[width=0.49\linewidth]{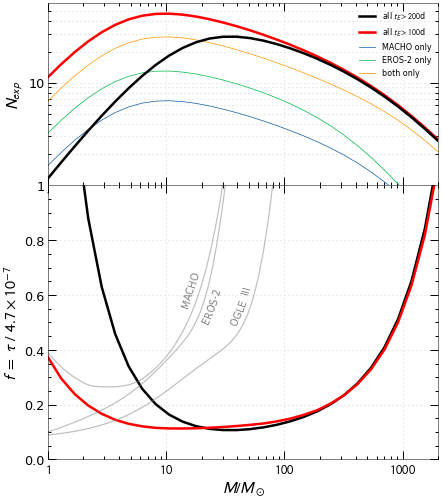}
    \caption{
Top : 
Number of events expected from a halo S-model entirely composed of compact objects of mass $M$:
blue (green) line, from source objects monitored only by MACHO (EROS-2);
orange line, from source objects monitored by both surveys; full red line shows the total; black line shows the total adding the constraint $t_E>200$ days in the analysis.
Bottom: $95\%$ CL upper limits on the fraction of the halo mass in the form of compact objects $f=\tau_{LMC}/4.7\times 10^{-7}$.
Limits obtained in this analysis are shown 
in red, and in black if we require $t_E>200$ days.  The grey curves correspond to the latest limits published by MACHO, EROS-2 and OGLE-III.
    }
    \label{fig:resultat}
\end{figure}

We finally subtract from our catalog the contamination by Galactic stars, estimated to be less than $5\%$ by
counting stars in the GAIA catalog\cite{2020Gaia_arXiv} in adjacent fields, and
we consider that $10\%$ of microlensing events may escape detection due to
lens binarity\cite{Tisserand_2007,Wyrzykowski_2011}.
The top panel of Fig. \ref{fig:resultat} shows, as a function of $M$, the expected number of detected
microlensing events for the halo S-model, assuming it consists entirely of $M$ mass deflectors,
using the efficiency shown in Fig. \ref{fig:tEdist_eff} for the simulation with blending.
For any lens mass distribution, the expected number of events is simply calculated by integrating the $N_{exp}(M)$ curve, weighted by the mass distribution.

We perform a Bayesian analysis taking into account the expected
$N_{exp}^{(s)}\sim 0.64$ disk and self-lensing events with $t_E>100\,$days.
The excluded halo fraction with 95\% CL, $f(M)$,
as a function of the deflector mass is shown in the lower panel of Fig. \ref{fig:resultat} by the red curve.  
We tested the robustness of this result by restricting our selection to events with
$t_E>200\,$days. By doing so, $N_{exp}^{(s)}$ becomes negligible ($<0.05$ event) and no candidates are retained.
The resulting exclusion limit, shown as the black curve,
is weaker at the lower mass end, but unchanged on the high mass side.

Finally, the combination of the limits obtained by EROS-1\cite{1997A&A...324L..69R}, EROS-2\cite{Tisserand_2007}, and the present analysis\cite{Blaineau2022} shows that deflectors with any mass distribution in the range $10^{-7}<M<200 M_\odot$ 
can not contribute more than $f \sim 20\%$ at 95\% CL to the halo mass, assuming a standard 
spherical halo.

\section{Conclusions and perspectives}
\label{section:discussion}
We conclude that massive compact objects with masses up to a thousand solar masses, similar to the ones observed by LIGO and VIRGO as binary BH mergers, do not make up a major fraction of the Milky Way dark matter, at least if assumed to be distributed as a standard spherical halo. Such BHs are more likely to be found in structures following the visible mass distribution, and could be searched for through microlensing toward the Galactic bulge and spiral arms, in the long $t_E$ tail of event duration distribution,
extending the previous searches and analysis\cite{Hamadache2006,Moniez2017,Mroz2020}.
\section*{Acknowledgments}
We thank
$\L$ucasz Wyrzykowski and Christopher Stubbs for providing useful information. This work was supported by the Paris Ile-de-France Region.\\
\section*{References}
\bibliography{citations_microlensing-combined}

\begin{thebibliography}{10}

\bibitem{Paczynski86}
B.~{Paczynski}.
\newblock {\em \apj}, 304:1, 1986.

\bibitem{Alcock2000}
{Alcock, C., et al. (MACHO Coll.)}.
\newblock {\em \apj}, 542(1):281--307, 2000.

\bibitem{Blaineau2022}
{Blaineau, T. et al. (EROS Coll.)}.
\newblock {\em arXiv:2202.13819}, 2022.

\bibitem{Tisserand_2007}
{Tisserand, P., et al., (EROS-2 Coll.)}.
\newblock {\em \aap}, 469(2):387--404, 2007.

\bibitem{MACHO_2001}
{Alcock, C., et al. (MACHO Coll.)}.
\newblock {\em \apjl}, 550(2):L169--L172, 2001.

\bibitem{Wyrzykowski_2011}
{Wyrzykowski, \L., et al. (OGLE Coll.)}.
\newblock {\em \mnras}, 416(4):2949--2961, 2011.

\bibitem{Mirhosseini2018}
A.~{Mirhosseini} and M.~{Moniez}.
\newblock {\em A\&A}, 618:L4, 2018.

\bibitem{these_Blaineau}
Tristan Blaineau.
\newblock PhD thesis, October 2021.

\bibitem{Alcock1999}
{Alcock, et al. (MACHO Coll.)}.
\newblock {\em \pasp}, 111(766):1539--1558, 1999.

\bibitem{2020Gaia_arXiv}
{Brown, A.~G.~A., et al. (Gaia Coll.)}.
\newblock {\em arXiv:2012.01533}, 2020.

\bibitem{1997A&A...324L..69R}
{Renault, C., et al., (EROS-2 Coll.)}.
\newblock {\em \aap}, 324:L69--L72, 1997.

\bibitem{Hamadache2006}
{Hamadache, C., et al. (EROS Coll.)}.
\newblock {\em \aap}, 454(1):185--199, 2006.

\bibitem{Moniez2017}
{Moniez, M. et al.}
\newblock {\em \aap}, 604:A124, 2017.

\bibitem{Mroz2020}
{Mróz, P. et al., (OGLE Coll.)}.
\newblock {\em The Astrophysical Journal Sup. Series}, 249(1):16, 2020.

\end{thebibliography}
\end{document}